\documentstyle[11pt,epsfig]{article}

\newlength{\dinwidth}
\newlength{\dinmargin}
\def\lapproxeq{\lower .7ex\hbox{$\;\stackrel{\textstyle<}{\sim}\;$}}
\def\gapproxeq{\lower .7ex\hbox{$\;\stackrel{\textstyle>}{\sim}\;$}}
\def\alps{\alpha_s}
\def\alpsb{\bar{\alpha}_s}
\def\gbar{\bar{\gamma}}

\begin{document}

\titlepage
\begin{flushright}
MC-TH-98/23 \\
December 1998
\end{flushright}

\begin{center}
{\Large \bf BFKL at next-to-leading order} \\
\vspace*{1cm}
J.R.~Forshaw \\
Department of Physics and Astronomy, \\
University of Manchester,\\
Manchester, M13 9PL, England.\\
\vspace*{0.4cm}
G.P.~Salam\\
INFN, Sezione di Milano,\\
Via Celoria 16,\\
20133 Milano, Italy.\\
\vspace*{0.4cm}
R.S.~Thorne \\
Theoretical Physics, Department of Physics, \\
Univerisity of Oxford,\\
Oxford, OX1 3NP, England.
\end{center}

\vspace*{0.4cm}
\begin{abstract}
This is a summary of the contributions on the next-to-leading order
corrections to the BFKL equation which were presented to the  
`Small-$x$ and Diffraction' working group at the 1998 Durham Workshop on 
HERA Physics. 
\end{abstract}

The original BFKL equation \cite{BFKL} allowed the computation of
elastic scattering amplitudes in the Regge limit to leading
logarithmic (LO) accuracy, i.e. all terms $\sim (\alps \ln s)^n$ are
summed. The key element is the gluon Green function:
\begin{equation}
F(s,k) = \int \frac{d \gamma}{2 \pi i} \frac{d \omega}{2 \pi i}
\left( \frac{s}{s_0} \right)^{\omega} \frac{1}{\omega-\alpsb \chi(\gamma)}
\left( \frac{k_1^2}{k_2^2} \right)^{\gamma}. \label{lo}
\end{equation}
The scale $s_0$ is arbitrary, $\alpsb = N_c \alps/\pi$ is fixed and, at 
leading order,
\begin{equation}
\chi(\gamma) = \chi_0(\gamma) = -2\gamma_E - \psi(\gamma)-\psi(1-\gamma).
\end{equation}
The $s \to \infty$ behaviour of the amplitude is governed by the
right-most $\omega$-plane singularity which occurs at $\gamma = \gbar$
where
\begin{equation}
\omega = \alpsb \chi(\gbar).
\end{equation}
The conformal invariance of the LO kernel allows us to write equation
(\ref{lo}).  The NLO corrections \cite{NLO} break the conformal
invariance by leading to the running of the QCD coupling. However, it
is possible to study the conformally invariant part of the NLO
equation. In this case, equation (\ref{lo}) is still relevant but now
the kernel is
\begin{equation} 
\chi(\gamma) = \chi_0(\gamma) + \alpsb \chi_1(\gamma).
\end{equation}
The NLO kernel depends upon the scale $s_0$ in a straightforward way
($s_0 = k_1 k_2$ is usually chosen).  The NLO corrections contained in
$\chi_1$ are huge for typical values of $\alps$ which is a cause for
serious concern since we are attempting a perturbative expansion
\cite{NLO,Ross,Levin}.  For example, at $\gamma=1/2$
$$\chi(\gamma) = \omega_0 (1 - 6.5 \ \alpsb)$$
where $\omega_0 =
\alpsb \chi_0(1/2) = 4 \ln2 \ \alpsb$ is the leading eigenvalue of the
LO BFKL kernel.

Marcello Ciafaloni and Gavin Salam pointed out that the problem is
alleviated by identifying and resumming a large part of the NLO
corrections: that with double logarithms in the transverse momenta
(corresponding to the $1/\gamma^3$ and $1/(1-\gamma)^3$ parts of the
NLO kernel).  After ensuring the consistency of these double
logarithms to all orders, the perturbation series is much more
convergent.

The double logarithms are closely associated with the choice of scale
$s_0$. Schematically one can see that they arise through a
change of scale $s_0$ from $k_1^2$ to $k_1 k_2$. In the situation with
$k_1\gg k_2$, the quantity which exponentiates is roughly
$$
  \alpsb \ln\left(\frac{s}{k_1^2}\right) 
         \ln\left(\frac{k_1^2}{k_2^2}\right)
 = \alpsb \ln\left(\frac{s}{k_1 k_2}\right) 
         \ln\left(\frac{k_1^2}{k_2^2}\right) -
\frac12 \alpsb \ln^2\left(\frac{k_1^2}{k_2^2}\right)\,.
$$
If one writes one's expansion in terms of scale $s_0=k_1k_2$ (i.e.\ 
the RHS) the exponentiation leads not just to a large NLO term, but
also to a whole series of NNLO, NNNLO, $\ldots$ terms with double
transverse logarithms, which can be large. These should therefore be
resummed. At first sight the simplest way to do so seems be to rewrite
one's expansion in terms of $s_0=k_1^2$ (LHS). The problem is that
such a scale choice spoils the convergence (again by double
logarithms) of the kernel in the case when $k_1 \ll k_2$, where the
`natural' (DGLAP) scale choice is instead $s_0 = k_2^2$.

A way of treating both regions on an equal footing while correctly
resumming the double logarithms was proposed in \cite{Salam} and is related
to an approach suggested by the Lund group \cite{LDC}. Figure 1 shows
that such a resummation greatly improves the convergence of the
kernel. The various schemes correspond to different treatments of less
singular terms (in particular those related to collinear branching and
the running coupling), which remain to be accounted for in such a
way as to ensure consistency with the renormalisation group (see
below, and also \cite{CC}).

\begin{figure}[htbp]
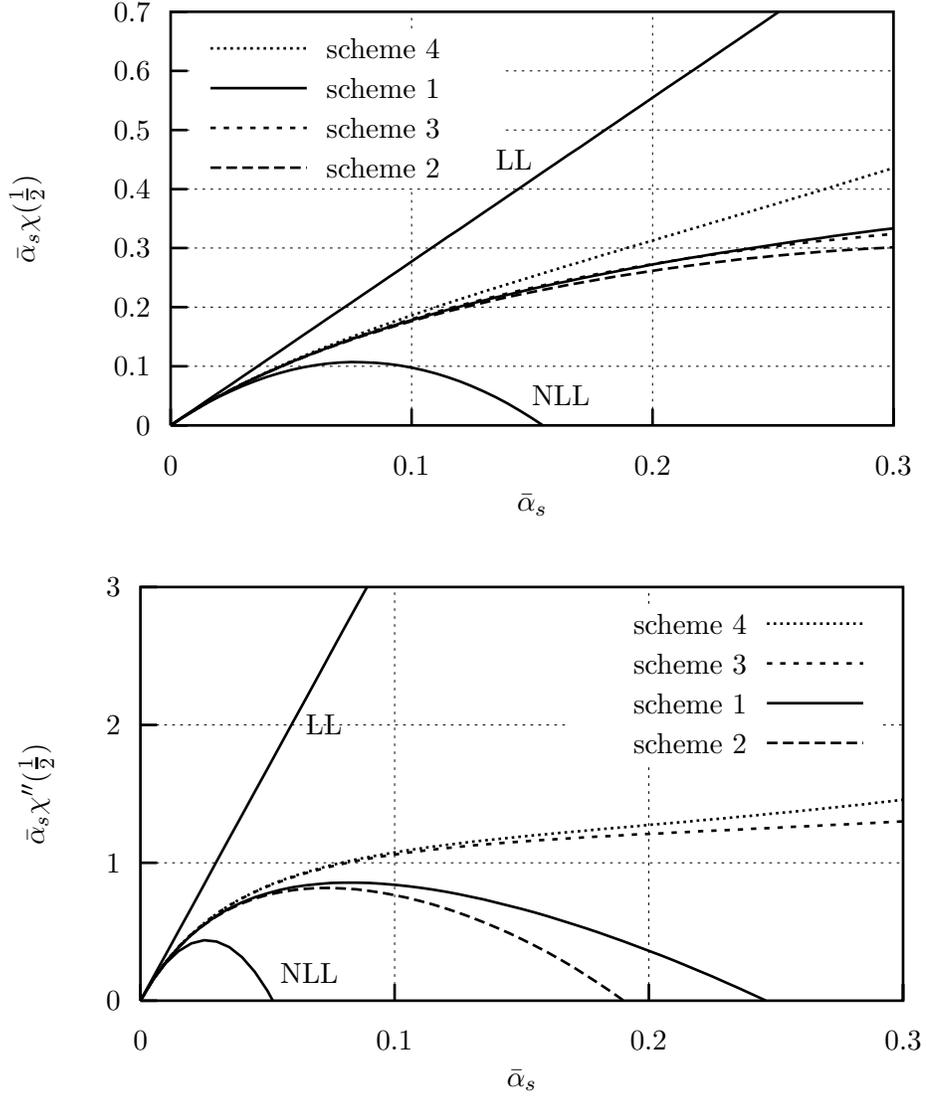

  \begin{center}
    \input{ltnl.pstex}
    \input{ltnl.deriv.pstex}
    \caption[]{The result of resummation on $\chi$ and its second
      derivative; shown for $n_f=0$. A symmetric version of the pure
      NLO kernel has been used to redefine $k^{2(\gamma-1)}$ as
        $\sqrt{\alps(k^2)}k^{2(\gamma-1)}$ in the eigenfunctions.}
  \end{center}
\end{figure}

The phenomenological importance of matching to the known DGLAP limit
and of including kinematical and other `finite $z$' effects, e.g.\
coherence, have been and will continue to be explored by a number of
groups. For example, see the contribution of Kharrazhia and Stasto in
these proceedings \cite{Hamid}. Lynn Orr showed us the crucial
importance of imposing the correct kinematics. She focussed on the
azimuthal decorrelation in dijet production at the Tevatron where the
data were compared with results from a BFKL Monte Carlo with the
correct kinematics imposed \cite{Orr}.

So far we ignored the fact that the QCD coupling runs. At NLO we need
to account for the breaking of conformal invariance due to running
coupling effects. The eigenfunctions of the kernel are no longer
simple powers of the momenta and a more sophisticated approach is
required. Since the BFKL formalism integrates over all momenta, the
running of the coupling invariably leads to divergences associated
with the failure of QCD perturbation theory in the infrared. To
regularise the theory, one might take a phenomenological approach and
introduce by hand some regulator, e.g.\ a gluon mass to freeze the QCD
coupling at some low scale. This has been done before, in the case of
the LO BFKL equation \cite{mass}. However, in single scale problems
like deep inelastic scattering at small $x$ one does not need to
speculate. The collinear factorisation of deep inelastic
cross-sections allows the ill behaved infrared dynamics to be
factorised into parton distribution functions and it is their scale
dependence that is predicted by renormalisation group (RG) (i.e.
DGLAP) equations of QCD. Ciafaloni stressed the importance of working
within a RG consistent framework.

A systematic approach to small-$x$ structure functions was presented by Robert Thorne.
A few years ago Catani showed how to write the DGLAP equations in a manifestly 
factorisation scheme independent way \cite{Catani}:
\begin{equation}
\frac{\partial F_i}{\partial \ln Q^2} = \Gamma_{ij} F_j
\end{equation}
where $i,j = 2,L$ and $\Gamma_{ij}$ is the physical anomalous dimension matrix. 
The NLO corrections to the BFKL equation can be used to determine the scale 
dependence of the structure functions $F_2$ and 
$F_L$.\footnote{Actually we don't have all the information yet since the 
coupling of the gluons to the photon via a quark loop has not yet been computed 
at NLO.} In particular, the physical anomalous dimension matrix is computed including
the complete two-loop DGLAP splitting functions supplemented with the leading twist
information coming from the NLO BFKL equation; corrections to this are 
strictly ${\cal O}(\alps(\mu))$. The only ambiguity in this procedure is 
in the choice of the renormalisation scale $\mu$. 
Thorne pointed out that the `natural' choice $\mu = Q$ has disasterous
consequences, as can be seen in the upper of Fig.2: the dotted line is 
the evolution of the longitudinal structure function at low $x$ with $\mu=Q$ 
at NLO, as compared to the LO evolution shown by the solid line
(in both cases $F_L(x,Q^2)$ is parameterized as $(0.1/x)^{-0.3}$ at 
$\ln(Q^2/\Lambda^2)=8$). Thorne adopted the BLM scale fixing procedure 
\cite{BLM} which, for some observable 
$$ R(Q) = A \alps(\mu) - \beta_0 \alps(\mu)^2 [ A \ln(Q/\mu)  + B ] + \alps(\mu)^2 C + \cdots, $$
states that the appropriate scale is determined by ensuring that
\begin{equation}
A \ln(Q/\mu) + B = 0.
\end{equation}
In this case $A$ and $B$ depend upon $\alps$ and $x$, and the scale choice
corresponds to
$$ \alps(\mu) \approx \frac{1}{\beta_0(\ln(Q^2/\Lambda^2)+3.6\alps(Q^2)
(\ln 1/x)^{1/2})}$$
at small $x$ and for all physical anomalous dimensions (and parton anomalous
dimensions in sensible schemes). 
This scale choice has the effect of absorbing the large higher order corrections into
the running of the coupling, i.e. subsequent corrections are small and support the
use of perturbation theory, as shown in the lower of Fig.2. 
It also has the peculiar effect of reducing the coupling
at low $x$. The physical origin of such strange behaviour was the cause 
for some discussion, though it now seems to be due to infrared diffusion in 
the BFKL equation only affecting the inputs for structure functions, while 
the evolution is sensitive to ultraviolet diffusion, and hence scales larger
than $Q^2$ as $x$ decreases \cite{Thorne}. Thorne also fixed the BLM scale for
$\Gamma_{22}$ ($\sim P_{qq}$) at high $x$, and showed us the result of fitting 
all the DIS data fitted by the MRST group using lowest order evolution
with the scale fixing. The resulting fit produces a remarkably small chi squared, as
can be seen in Table 1 which compares the latest MRST fit with the lowest order BLM
fit (LO(x)). It also leads to $F_L(x,Q^2)$ following the shape of $F_2(x,Q^2)$
at low $Q^2$, unlike the  conventional approaches.  

\begin{figure} 
\centerline{\epsfig{file=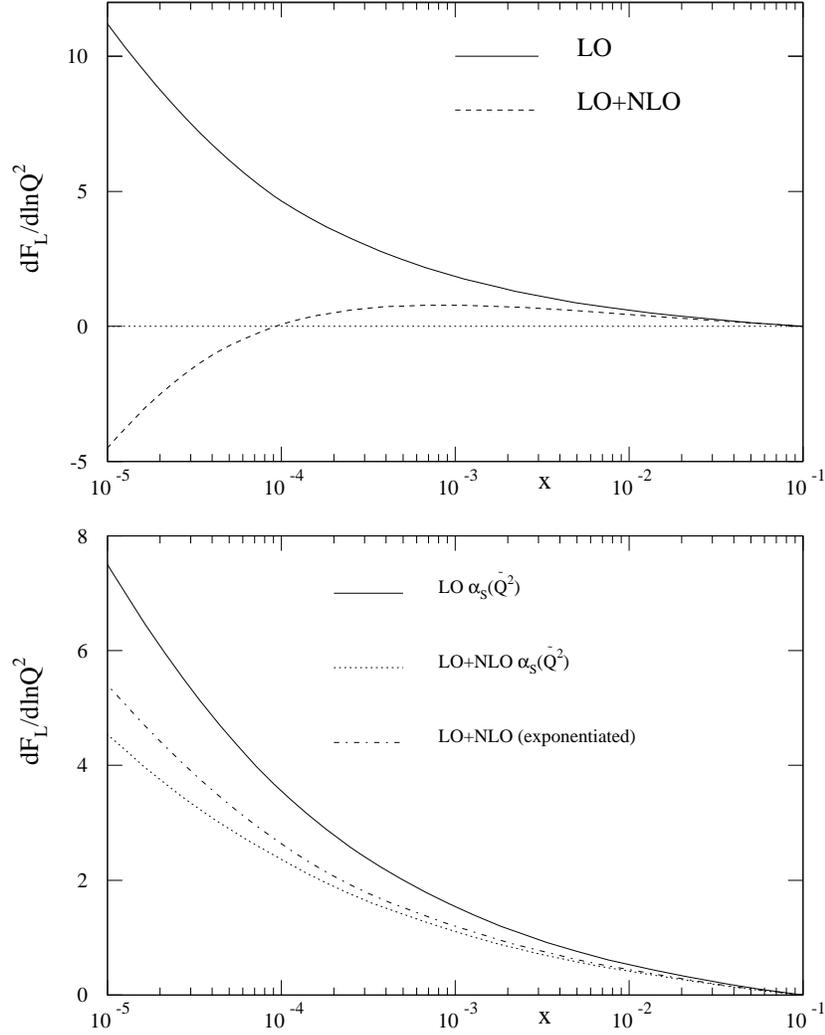,height=15cm,angle=0}}
\caption{Evolution of $F_L$ at LO and NLO.}
\end{figure}

\vspace{0.5cm}                                                              
\begin{center}                                                              
\begin{tabular}{|cccc|}\hline                                                              
Data set          & No. of   & LO(x) & MRST \\                                                              
                  & data pts &      &      \\ \hline                                                              
H1 $ep$           & 221      & 149  & 164      \\                                                              
ZEUS $ep$         & 204      & 246  & 270      \\                                                              
BCDMS $\mu p$     & 174      & 241  & 249      \\                                                              
NMC $\mu p$       & 130      & 118  & 141      \\                                                              
NMC $\mu d$       & 130      & 81   & 101      \\                                                              
SLAC $ep$         & 70       & 87   & 119      \\                                                              
E665 $\mu p$      & 53       & 59   & 58       \\                                                              
E665 $\mu d$      & 53       & 61   & 61       \\                                                              
CCFR $F_2^{\nu N}$& 66       & 57   & 93      \\                                                              
CCFR $F_3^{\nu N}$& 66       & 65   & 68       \\                                                              
NMC \it{n/p}      & 163      & 176  & 187      \\ \hline
 
total             & 1330     & 1339 & 1511     \\ \hline                                                  
\end{tabular}                                                              
\end{center}                                                              
\noindent Table 1: Comparison of $\chi^2$ for fits using the full
leading order (including $\ln(1/x)$ terms) renormalization scheme
consistent expression with BLM scale, and the two-loop MRST fit.

\end{document}